\documentclass[prc,aps,twocolumn,showpacs,preprintnumbers,eqsecnum,amsmath,amssymb,twocolomn]{revtex4-1}
\usepackage{graphicx}
\usepackage{epsfig}
\usepackage{dcolumn}
\usepackage{bm}

\begin{document}
\title{Nuclear pygmy modes and the dynamics of the nuclear skin}

\author{
 N.~Tsoneva$^{1,2}$, H.~Lenske$^{1}$}
\affiliation{
  $^1$Institut f\"ur Theoretische Physik, Universit\"at Gie\ss en,
  Heinrich-Buff-Ring 16, D-35392 Gie\ss en, Germany \\
$^2${Institute for Nuclear Research and Nuclear Energy, 1784 Sofia, Bulgaria}}
\begin{abstract}
The information on pygmy resonances reveals new aspects on the isospin dynamics of the nucleus with important astrophysical consequences. 
In this connection, the precise knowledge of nuclear response functions plays a key role in the determination of photonuclear reactions cross sections which are of importance for the synthesis of heavy neutron-rich elements.
For that purpose, a theoretical method based on density functional theory and multi-phonon approach is applied for investigations of nuclear excitations with different multipolarities and energies in stable and exotic nuclei. The possible relation of low-energy modes to the properties of neutron or proton skins is systematically investigated for isotonic and isotopic chains. 
Our studies of dipole and quadrupole response functions and the corresponding transition densities indicate new pygmy dipole and  pygmy quadrupole resonances, describing oscillations of the nuclear skin. 
Also, the presence of skins is found to affect the magnetic response of nuclei.
\end{abstract}

\maketitle

\section{Introduction}
The studies of nuclear electromagnetic response along isotonic and isotopic chains of nuclei led to surprising new observations of new modes
of excitations different from known surface vibrations and giant resonances
in nuclei close to the isospin N=Z symmetry. A clear example is the existence of an enhanced dipole strength located below or closely above the neutron particle emission threshold which was found to be a common feature of stable and unstable nuclei with neutron excess \cite{Zil02,Adr05,Vol06,Sch08,Ton10,Endr10}. 
As far as this bunching of dipole states with very similar spectroscopic features resemble a resonance, the new mode was named pygmy dipole resonance (PDR). Even though, the nature of the PDR is a subject of outstanding discussions, the most common theoretical interpretation associated it     
with oscillations of a weakly bound neutrons (for N/Z$>$1) or protons (for N/Z$\approx$1) with respect to an isospin symmetric core \cite{Vol06,Sch08,Ton10,Endr10,Tso04,Tso08,Paa07}. 
A question, coming up immediately in this connection, is to what extent the presence of a neutron or proton skin will affect excitations of other multipolarities and parities and {\em vice versa}. Promising candidates are low-energy 1$^{+}$, 2$^{+}$, 3$^{-}$ states ect., especially those in excess of the spectral distributions known from stable nuclei. Recently, quadrupole response functions were investigated, theoretically in neutron-rich Sn nuclei. A close connection of low-energy 2$^+$ excitations and nuclear skins was found. These quadrupole states were related to pygmy quadrupole resonance (PQR) \cite{Tso11}. 
In this connection, QRPA and QPM calculations of dipole and quadrupole strength distributions for pygmy resonances in nuclei with Z=12,13,50 and N=50,82 are discussed.  

\section{The theoretical model}

The model Hamiltonian resembles in structure the standard quasiparticle-phonon model (QPM) \cite{Sol76} but in detail differs in the physical content in important aspects as discussed in \cite{Tso04,Tso08}:

\begin{equation}
{H=H_{MF}+H_M^{ph}+H_{SM}^{ph}+H_M^{pp}} \quad .\label{hh}
\end{equation}
Here, $H_{MF}=H_{sp}+H_{pair}$ is the mean-field part defining the single particle properties including pairing interactions for protons and neutrons. Different from the standard QPM scheme this part is obtained self-consistently by a fully microscopic Hartree-Fock-Bogoliubov (HFB) approach \cite{Hofmann}.  
$H_M^{ph}$, $H_{SM}^{ph}$ and $H_M^{pp}$ are residual interactions,
taken as a sum of isoscalar and isovector separable multipole and
spin-multipole interactions in the particle-hole ($ph$) and multipole
pairing interaction in the particle-particle $(pp)$ channels. The model parameters are fixed either empirically \cite{Vdo} or by reference to QRPA calculation performed within the density matrix expansion (DME) of G-matrix interaction discussed in Ref. \cite{Hofmann}.

\subsection{The nuclear ground state}
The reliable description of ground state properties is of genuine importance for extrapolations of QRPA and QPM calculations into unknown mass regions. Taking advantage of the
Kohn-Sham theorem \cite{HoKohn:64,KohnSham:65} of density functional theory (DFT) the total
binding energy $B(A)$ of the nucleus could be expressed as an integral over an energy
density functional with (quantal) kinetic ($\tau$) and self-energy
parts, respectively,
\begin{equation}
B(A)=\int{d^3r\left( \tau(\rho)+\frac{1}{2}\rho
U(\rho)\right)}+E_{pair}
\end{equation}
\[
=\sum_j{v_j^2\left(e_j-<\Sigma>_j+\frac{1}{2}<U>_j\right)}+E_{pair}
\quad ,
\]
and pairing contributions are indicated by $E_{pair}$. The second
relation is obtained from HFB theory with 
occupancies $v^2_j$ and
potential energies $<U>_j$ of the occupied levels $j$, see e.g.
\cite{Hofmann}. Above, $U(\rho)$ is the proper self-energy,
i.e. not including the rearrangement contributions from the
intrinsic density dependence of nuclear interactions
\cite{Hofmann}. Hence, $U(\rho)$ has to be distinguished
from the effective self-energy obtained by variation
\begin{equation}
\Sigma(\rho)=\frac{1}{2}\frac{\partial \rho \rho U(\rho)}{\partial
\rho}
\end{equation}
and appearing in the single particle Schroedinger equation. In
order to keep the QPM calculations feasible we choose $\Sigma
\equiv U_{WS}$ to be of Wood-Saxon (WS) shape with adjustable
parameters. By inversion and observing that the densities and
potentials in a finite nucleus are naturally given parametrically
as functions of the radius $r$, we find
\begin{equation}
\rho(r) U(r)= -2 \int_r^\infty{ds \frac{\partial \rho(s)}{\partial
s} U_{WS}(s)} \quad .
\end{equation}
Evaluating these relations with the microscopic proton and neutron
densities obtained by solving the Schroedinger equation with
$U_{WS}$ the potential $U(\rho)$ is the self-consistently derived
reduced self-energy entering e.g. into the binding energy.

\begin{figure}
\begin{center}
\resizebox{1.10\columnwidth}{!}{
\includegraphics{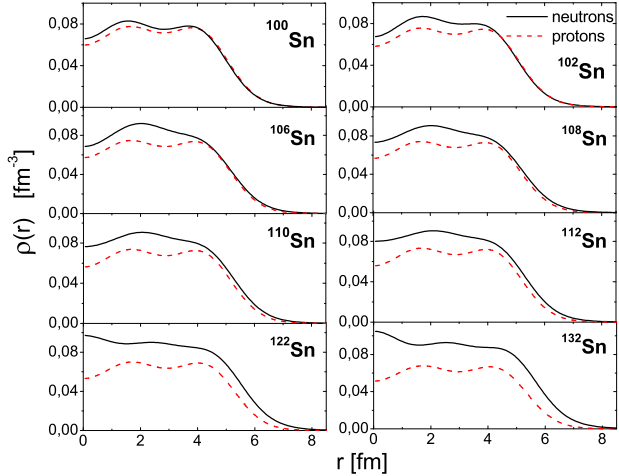}
}
\caption[]{BCS ground state densities of Sn isotopes obtained by the
phenomenological DFT approach and used in the QPM
calculations.}
\label{FIG1}
\end{center}
\end{figure}

\begin{figure}
\begin{center}
\resizebox{1.10\columnwidth}{!}{
\includegraphics{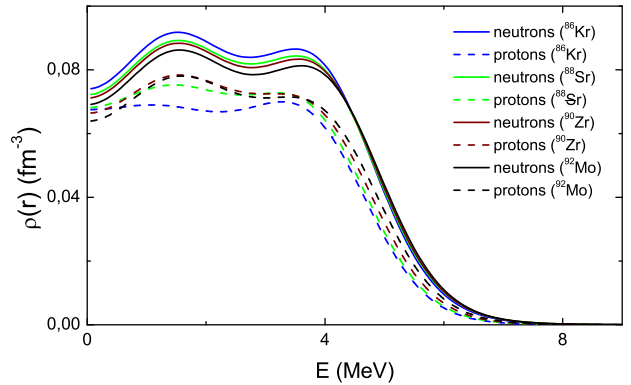}}
\caption[]{BCS ground state densities in N=50 obtained by the
phenomenological DFT approach and used in the QPM
calculations.}
\label{FIG2}
\end{center}
\end{figure}

In practice, for a given nucleus of mass $A$ the depth of the
central and spin-orbit potentials, radius and diffusivity
parameters of $U_{WS}$ are adjusted separately for protons and
neutrons to the corresponding single particle separation energies,
the total binding energy \cite{Audi95}, the charge radii and
(relative) differences of proton and neutron root-mean-square
(RMS) radii,
\begin{equation}\label{dr}
\delta r=\sqrt{<r^2>_n}-\sqrt{<r^2>_p} \quad ,
\end{equation}
from our previous HFB calculations \cite{Hofmann}.
In Ref. \cite{Tso08} theoretically obtained RMS radii are compared to those determined from charge exchange reactions for a number of Sn isotopes. 
The approach sketched above leads to very
satisfactory results on binding energies and proton-neutron
RMS-differences as shown in \cite{Tso04,Tso08}. A smooth dependence of the parameters on $A$ is found which supports the reliability of the method.

Calculations of ground state neutron and proton densities for Z=50 and N=50 are shown in Fig. \ref{FIG1} and Fig. \ref{FIG2}. Results from our calculations in N=82 isotones could be found in Ref. \cite{Tso08}. Of special importance are the nuclear surface regions, where the formation of a skin takes place. A common observation
found in the investigated isotopic and isotonic chains of nuclei is that the thickness of the neutron skin is related to the neutron-to-proton ratio N/Z.  Thus, in Sn isotopes with A$\geq$106 the neutron distributions begin to extend beyond the proton density and the effect continues to
increase with the neutron excess, up to $^{132}Sn$. Thus, these
nuclei have a neutron skin. The situation reverses in $^{100-102}Sn$,
where a tiny proton skin appears. 
For the case of N=50 isotones the neutron skin thickness decreases from $^{86}$Kr toward $^{92}$Mo which has the largest proton number Z. 

\subsection{The nuclear excited states}
The nuclear excitations are expressed in terms of quasiparticle-random-phase- approximation (QRPA) phonons \cite{Sol76}:
\begin{equation}
Q^{+}_{\lambda \mu i}=\frac{1}{2}{
\sum_{jj'}{ \left(\psi_{jj'}^{\lambda i}A^+_{\lambda\mu}(jj')
-\varphi_{jj'}^{\lambda i}\widetilde{A}_{\lambda\mu}(jj')
\right)}},
\label{phonon}
\end{equation}
where $j\equiv{(nljm\tau)}$ is a single-particle proton or neutron state;
${A}^+_{\lambda \mu}$ and $\widetilde{A}_{\lambda \mu}$ are
time-forward and time-backward operators, coupling 
two-quasiparticle (2QP) creation or annihilation operators to a total
angular momentum $\lambda$ with projection $\mu$.
The excitation energies of the phonons and the time-forward and time-backward amplitudes
$\psi_{j_1j_2}^{\lambda i}$ and $\varphi_{j_1j_2}^{\lambda i}$ in Eq.~(\ref{phonon}) are determined by solving QRPA equations \cite{Sol76}.

Furthermore, the QPM provides a microscopic approach to multiconfiguration mixing \cite{Sol76}. 
For spherical even-even nuclei the model Hamiltonian is diagonalized on an orthonormal set of wave functions constructed from one-, two- and three-phonon configurations \cite{Gri94}. 

\begin{equation}
\Psi_{\nu} (JM) =
 \left\{ \sum_i R_i(J\nu) Q^{+}_{JMi}
\right.
\label{wf}
\end{equation}
\[
\left.
+ \sum_{\scriptstyle \lambda_1 i_1 \atop \scriptstyle \lambda_2 i_2}
P_{\lambda_2 i_2}^{\lambda_1 i_1}(J \nu)
\left[ Q^{+}_{\lambda_1 \mu_1 i_1} \times Q^{+}_{\lambda_2 \mu_2 i_2}
\right]_{JM}
{+ \sum_{_{ \lambda_1 i_1 \lambda_2 i_2 \atop
 \lambda_3 i_3 I}}}
\right.
\]
\[
\left.
{T_{\lambda_3 i_3}^{\lambda_1 i_1 \lambda_2 i_2I}(J\nu )
\left[ \left[ Q^{+}_{\lambda_1 \mu_1 i_1} \otimes Q^{+}_{\lambda_2 \mu_2
i_2} \right]_{IK}
\otimes Q^{+}_{\lambda_3 \mu_3 i_3}\right]}_{JM}\right\}\Psi_0
\]
where R, P and T are unknown amplitudes, and $\nu$ labels the
number of the  excited states.

The electromagnetic transition matrix elements are calculated for transition operators including the interaction of quasiparticles and phonons \cite{Pon98}
where exact commutation relations are implemented which is a necessary condition in order to satisfy the Pauli principle.

The QPM allows for sufficiently large configuration spaces such that a unified description of low-energy single-  and multi-particle states is achieved also for the GDR energy regions. Such a unified treatment is exactly what is required in order to separate the multi-phonon and the genuine PDR $1^{-}$
strengths in a meaningful way.

\section{Discussion}

\begin{figure*}
\begin{center}
\resizebox{2.0\columnwidth}{!}{
\includegraphics{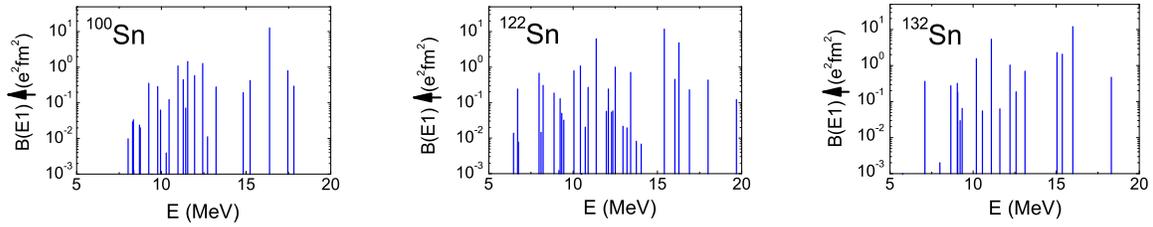}}\caption{QRPA calculations of electric dipole strength distributions in Sn isotopes.}
\label{FIG3}
\end{center}
\end{figure*}

\begin{figure*}
\begin{center}
\resizebox{2.0\columnwidth}{!}{
\includegraphics{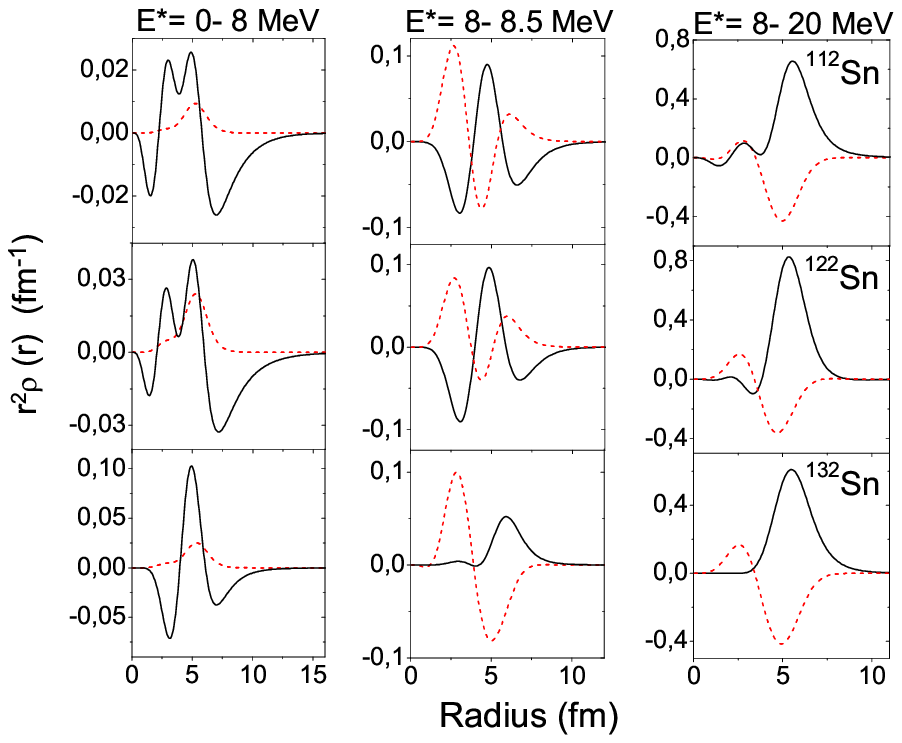}
\includegraphics{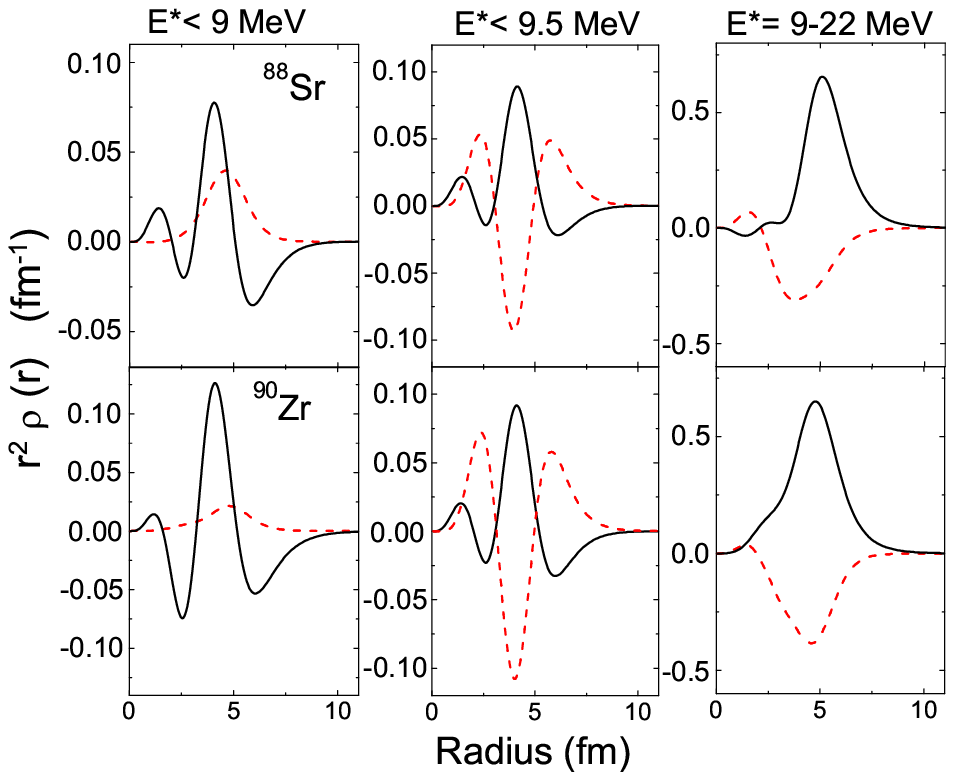}}
\caption{Dipole one-phonon proton and neutron transition densities in (left panel) Z=50 isotopes and (right panel) N=50 isotones.}
\label{FIG4AB}
\end{center}
\end{figure*}

\begin{figure*}
\begin{center}
\resizebox{2.0\columnwidth}{!}{
\includegraphics{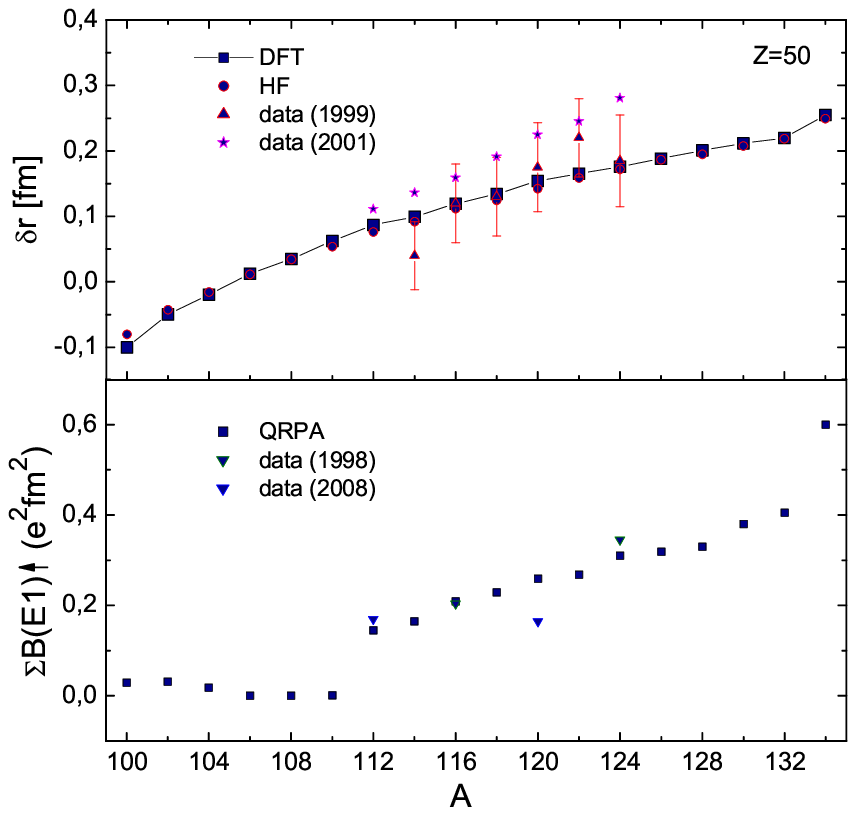}
\includegraphics{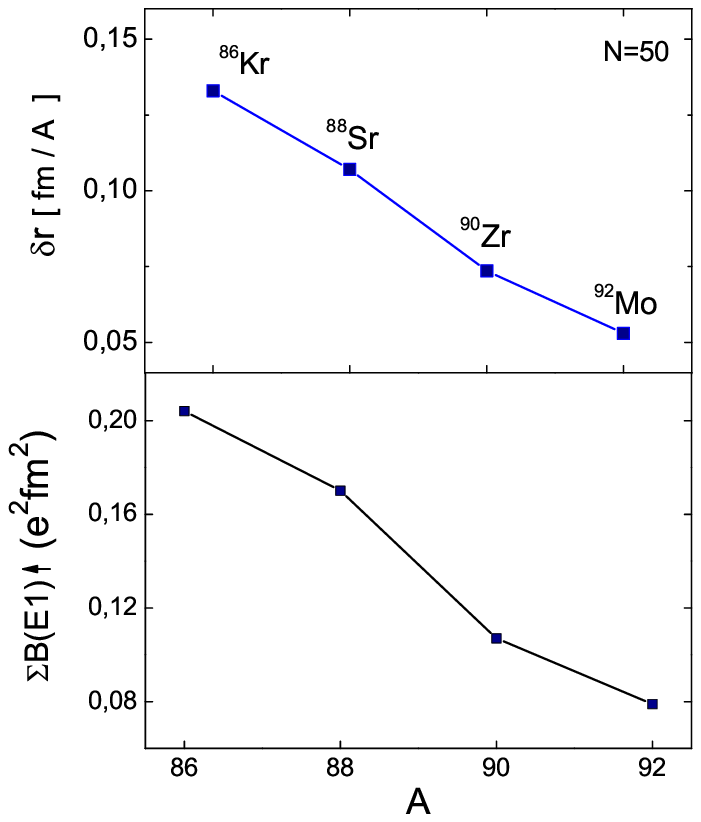}}
\caption[]{(left panel) QRPA results for the total PDR strengths in the
$^{100-134}$Sn isotopes are displayed for comparison
together with the nuclear skin thickness $\delta r$, eq.
\protect(\ref{dr}). Experimental data on the
total PDR strengths in $^{112,116,120,124}$Sn [19,20]) are also shown. In
the lower panel, the skin thickness derived from charge exchange
reactions by Krasznahorkay et al. published in Refs. [21] are indicated.
(right panel) QRPA results for the total PDR strengths in N=50 isotones.}
\label{FIG5AB}
\end{center}
\end{figure*}

\begin{figure*}
\begin{center}
\resizebox{2.0\columnwidth}{!}{
\includegraphics{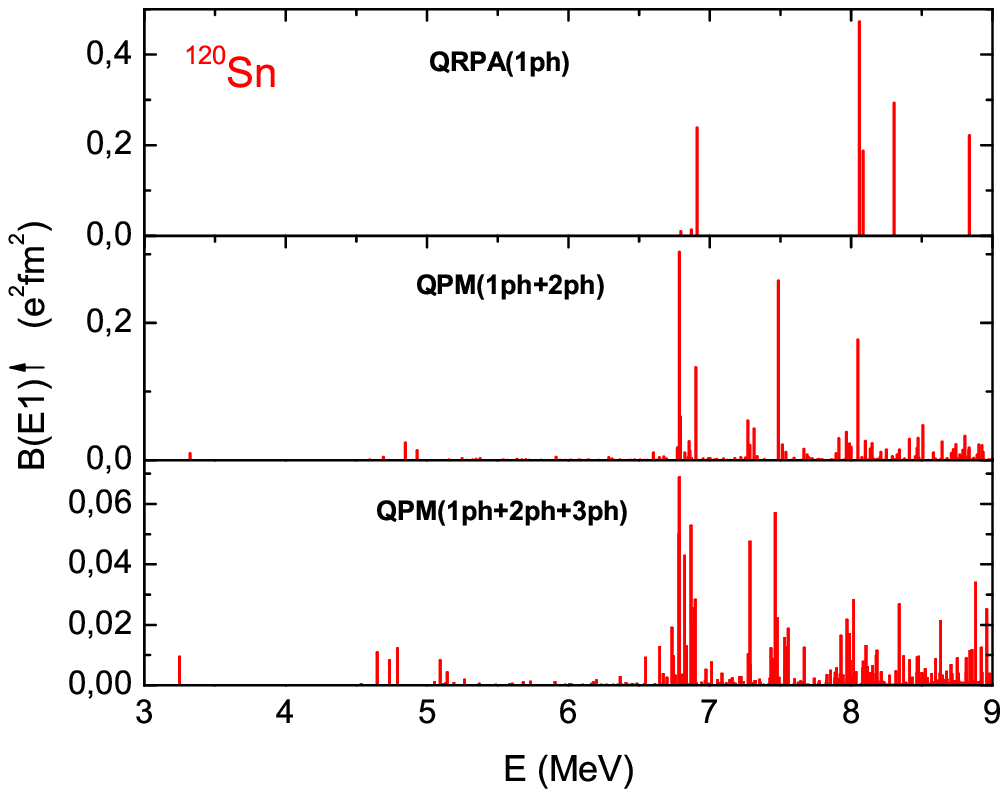}
\includegraphics{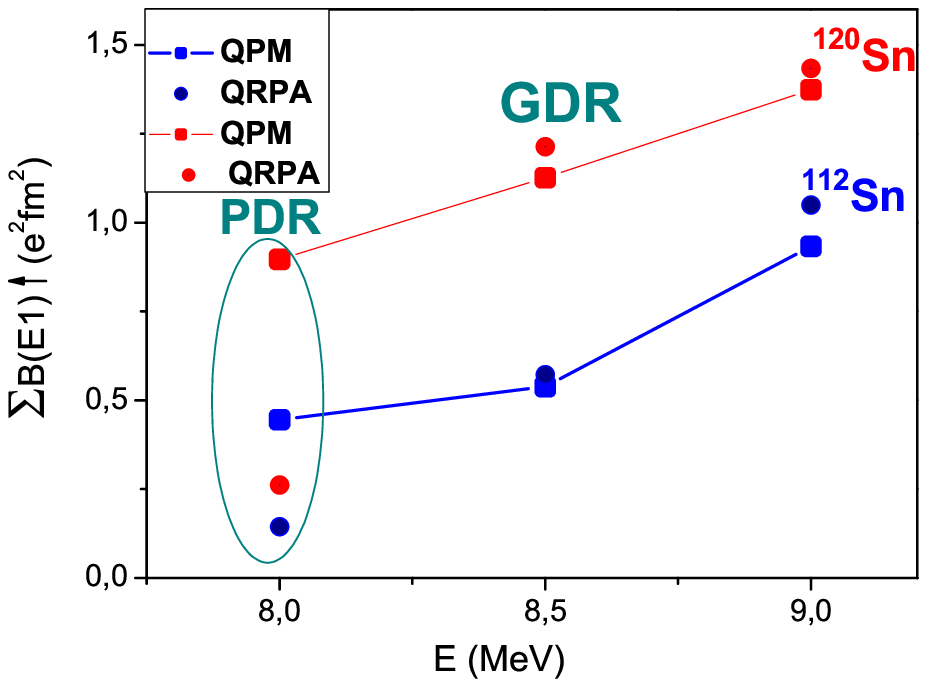}}
\caption{(left panel) Calculations of electric dipole strength up to E$^{*}$= 9 MeV in $^{120}$Sn: a/QRPA, b/ two-phonon QPM and c/ three-phonon QPM;
(right panel) Comparison of QRPA and three-phonon QPM B(E1) strengths summed up to E$^{*}$= 8 MeV (PDR region), 8.5 MeV (GDR region) and 9 MeV (GDR region), respectively.}
\label{FIG6AB}
\end{center}
\end{figure*}

\begin{figure*}
\begin{center}
\resizebox{2.10\columnwidth}{!}{
\includegraphics{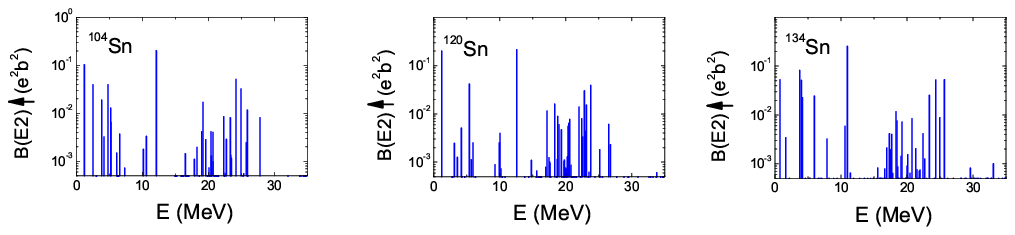}}
\caption{QRPA calculations of electric quadrupole strength distributions in Sn isotopes.\label{FIG7}}
\end{center}
\end{figure*}

\begin{figure*}
\begin{center}
\resizebox{2.0\columnwidth}{!}{
\includegraphics{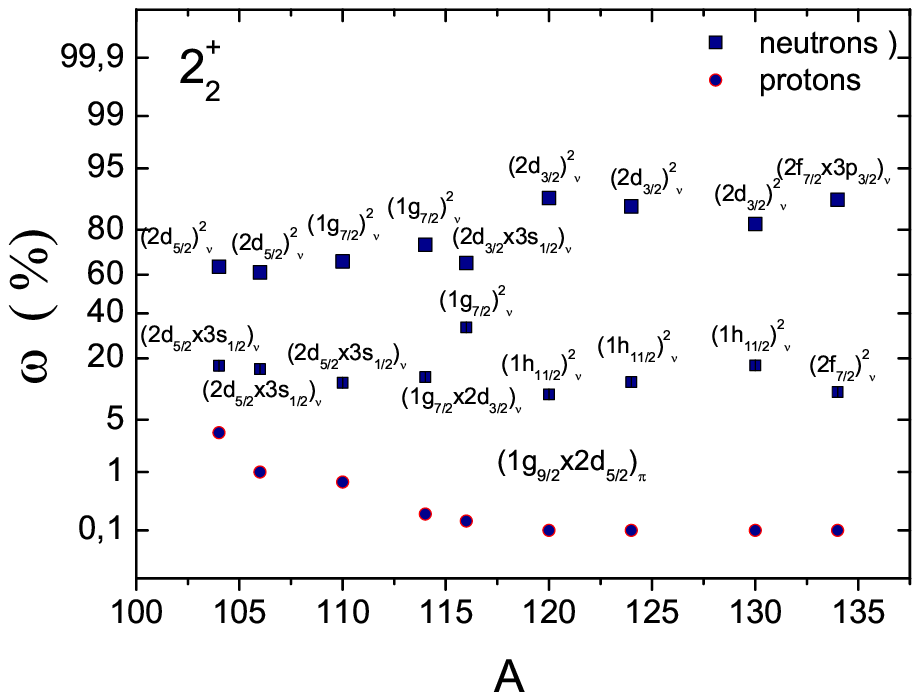}
\includegraphics{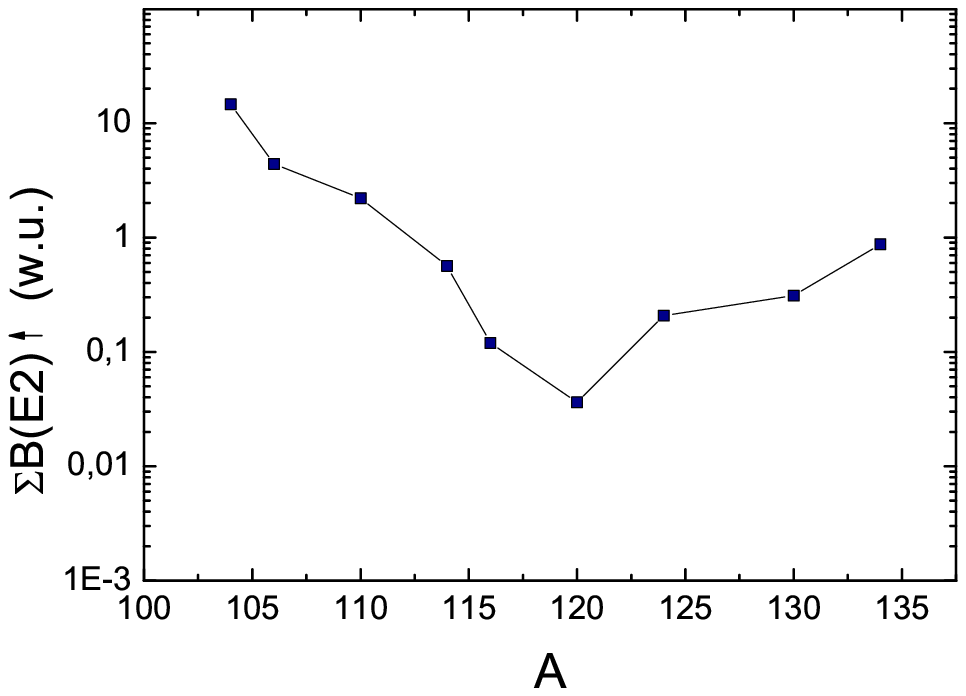}}
\caption{(left) QRPA structure of the $[2^{+}_{2}]_{QRPA}$ states calculated in Sn isotopes. The main neutron and proton 2QP components with a state vector probability $\omega_{j_{1}j_{2}}({\lambda i})$=$|\psi_{j_1j_2}^{\lambda
i}|^2-|\varphi_{j_1j_2}^{\lambda\mu i}|^2$ larger than 0.1$\%$ are presented;
(right) QRPA results for B(E2)$\uparrow$ transition probabilities  summed over the $[2^{+}_{2}]_{QRPA}$ and $[2^{+}_{3}]_{QRPA}$ states in Sn isotopes.}
\label{FIG8AB}
\end{center}
\end{figure*}

\begin{figure*}
\begin{center}
\resizebox{2.0\columnwidth}{!}{
\includegraphics{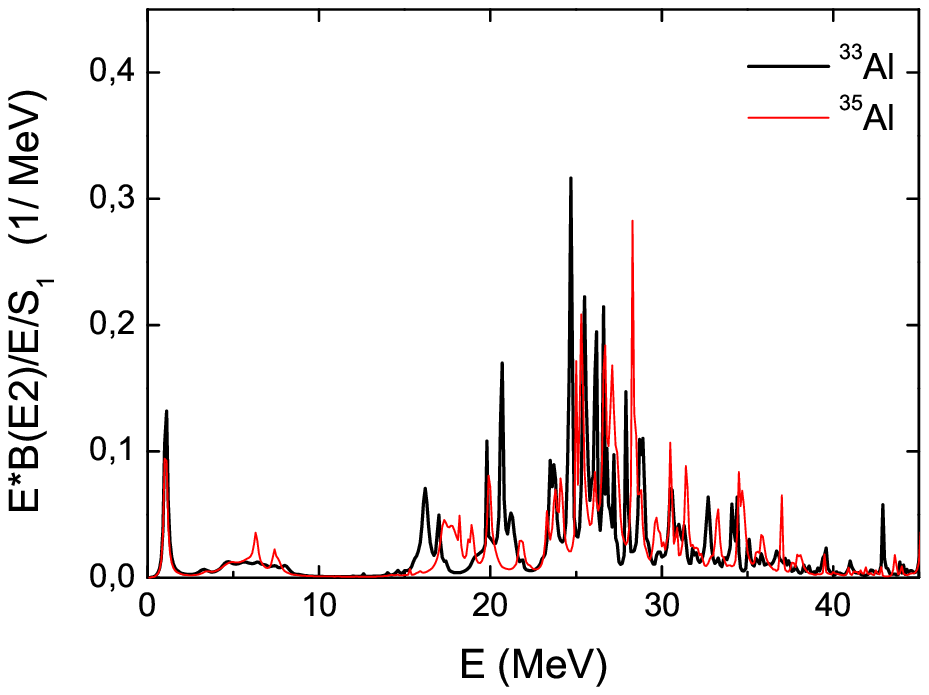}
\includegraphics{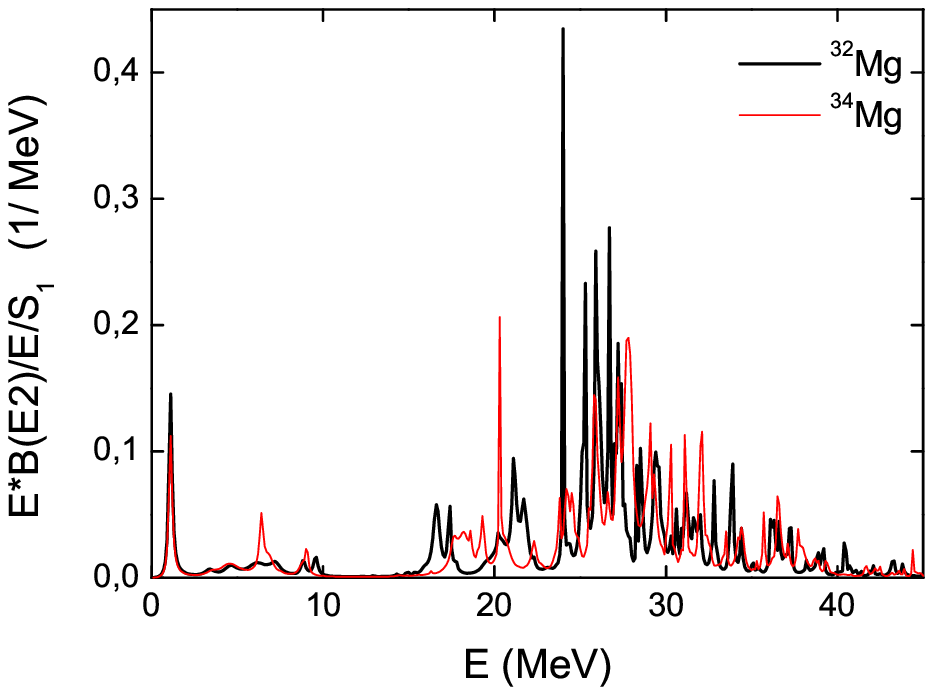}}
\caption{ QRPA calculations of B(E2) spectral distributions normalized to EWSR: $S_n=\Sigma_i E_i^n B(E2)$ in (left panel) $^{33}$Al and $^{35}$Al and (right panel) $^{32}$Mg and $^{34}$Mg.}
\label{FIG9AB}
\end{center}
\end{figure*}

\begin{figure*}
\begin{center}
\resizebox{2.0\columnwidth}{!}{
\includegraphics{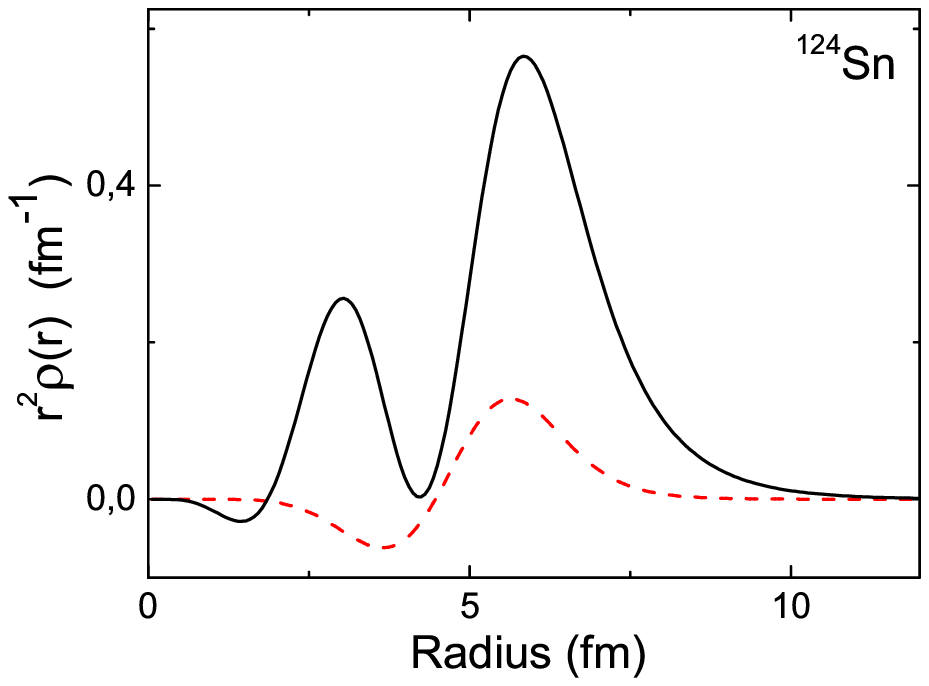}
\includegraphics{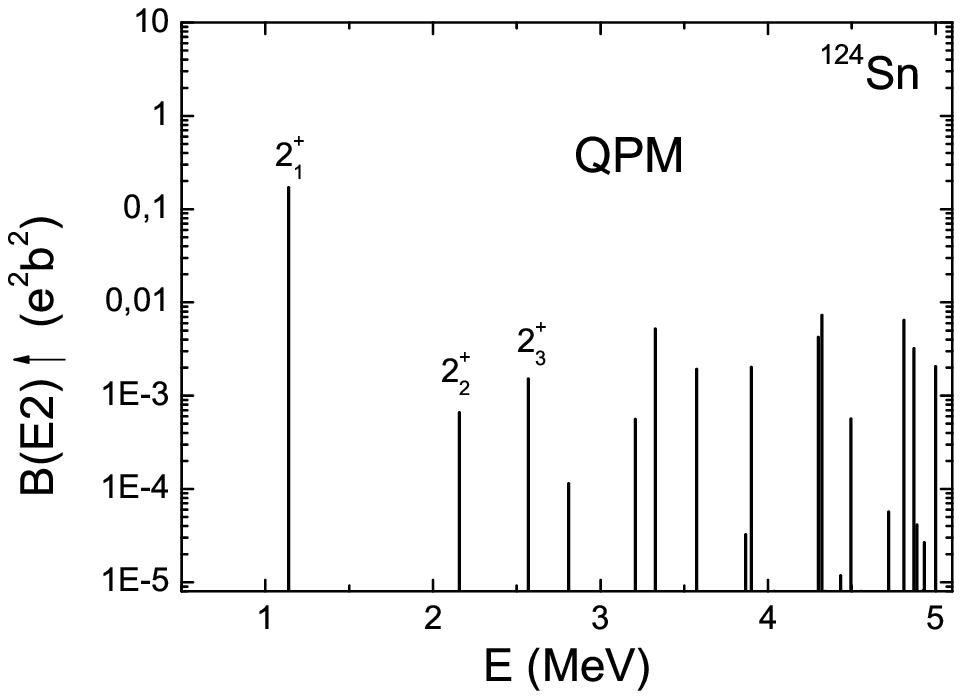}}
\caption{(left panel)QRPA calculations of quadrupole transition densities related to the main proton and neutron components in the state vectors of $[2^+_i]_{QRPA}$ states with excitation energies E$^{*}$= 2-5 MeV in $^{124}$Sn;
(right panel) Multiphonon calculations of B(E2) transitions probabilities of low-energy 2$^+$ excitations in $^{124}$Sn. 
\label{FIG10AB}}
\end{center}
\end{figure*}

\begin{figure*}
\begin{center}
\resizebox{2.1\columnwidth}{!}{
\includegraphics{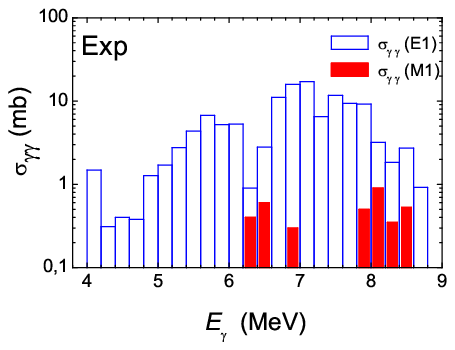}
\includegraphics{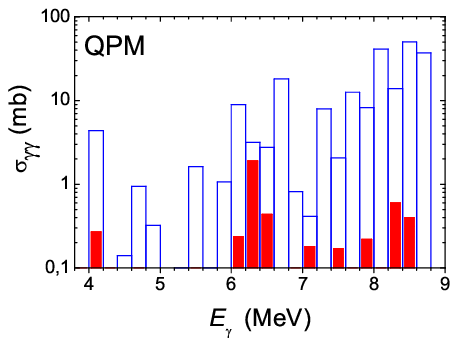}}
\caption{Experimental data (left panel) and QPM calculations 
(right panel) for $E$1 (electric dipole) and $M$1 (magnetic dipole) photoabsorption cross sections in $^{138}$Ba below 9.0 MeV. The cross sections are averaged over 0.2 MeV energy bins (see in Ref. 5 as well).
\label{FIG11AB}}
\end{center}
\end{figure*}

From QRPA calculations in $^{110-132}$Sn, a sequence of low-lying
one-phonon 1$^{-}$ states of almost pure neutron structure at excitation energies $E^*= 6- 7.5$~MeV is observed. These states correspond to excitations of least bound neutrons from the 3s-, 2p- and 2d-subshells with only a minor proton contribution of less than 1\%. 
Similar results are found also in N=50, 82 isotones \cite{Vol06,Sch08}. These low-energy dipole states have been related to neutron PDR.

In the lightest $^{100-104}$Sn isotopes, the lowest dipole excitations at E$^*$= 8.1-8.3 MeV are dominated by proton excitations involving
quasibound $2p_{3/2}$ and $1g_{9/2}$ proton states \cite{Tso08}.
Electromagnetic breaking of isospin symmetry is the main reason
for the persisting of low-energy dipole strength close to $^{100}$Sn.
QRPA calculations of dipole response functions in several tin isotopes are presented in Fig. \ref{FIG3}. The observed sequence of low-energy neutron or proton dipole states have been related to neutron or proton PDR, respectively \cite{Tso08}.
For a more detailed insight into the characteristic features of the
dipole excitations we consider the evolution of the proton and
neutron transition densities in various energy regions. In
Fig. \ref{FIG4AB} (left panel) we display dipole QRPA transition
densities for several tin isotopes
for three different regions of excitation
energies: the low-energy PDR region below the neutron emission
threshold, the transitional region up to the GDR and in the GDR
region and beyond. The
transition densities displayed in Fig. \ref{FIG4AB} (left panel) were obtained by
summing over the transition densities of the individual one-phonon 1$^-$ states located in the energy intervals denoted at the top of each column of the figures.
A common features of the presented $^{112,122,132}$Sn cases is that up to $E^*= 8 $~MeV the protons and neutrons oscillate in phase in the nuclear interior,
while at the surface only neutron transitions contribute. This
pattern is generic to the lowest dipole states making it meaningful
to distinguish these excitations from the well known GDR states.
Hence, we are allowed to identify the PDR as a distinct and unique excitation different from the giant dipole resonance (GDR). Similar results for the QRPA neutron and proton dipole transitions densities are obtained also in N=50 isotopes which are presented in Fig. \ref{FIG4AB} (right panel). The exited QRPA 1$^-$ states located below excitation energy E$^{*}<$9 MeV are related to PDR. 

The transition of the neutron PDR to a proton PDR manifests itself via the proton and neutron transition densities as well. Thus, in $^{100}$Sn the proton oscillations dominate at the nuclear surface due to the formation of a proton skin (see in Ref. \cite{Tso08}). In comparison, the presence of a neutron skin explains the neutron oscillations at the surface of tin nuclei with A$\le$106 shown in Fig. \ref{FIG1} (left panel).

In Fig. \ref{FIG5AB} the total PDR strength calculated in Z=50 (left panel) and 
N=50 (right panel) nuclei is presented in comparison with the 
neutron skin thickness $\delta r$, eq. (\ref{dr}) of these nuclides. It is shown that the two quantities are closely related. Here, the B(E1) sums are taken over QRPA dipole states associated with PDR according the analysis of transition densities and the structure of the state's vectors, respectively (see in Ref. \cite{Tso08} for more details). 
The close relationship of the PDR strength and the skin thickness is also confirmed in our investigations of N=82 nuclei \cite{Vol06,Ton10}.

The comparison of QRPA and two- and three-phonon QPM calculations of dipole strengths indicates that for the PDR region the coupling of QRPA PDR and GDR phonons and multiphonon states is very important. The result is a shift of E1 strength toward lower energies which is well visible in Fig. \ref{FIG6AB} (left panel). Consequently in the energy interval $0<E<8$ MeV the QPM calculations give about twice as much total E1-strength than the QRPA calculations, as shown in Fig. \ref{FIG6AB} (right panel). 
Differently, with the increase of the excitation energy toward the GDR the QRPA and QPM summed B(E1) transition strengths are of comparable size which suggests the prevalence of $p-h$ excitations of GDR type.
Thus, our QPM calculations of $^{112}$Sn (N/Z=1.24) and $^{120}$Sn (N/Z=1.4) confirm experimental and QRPA observations of increasing PDR strength with the increase of N/Z ratios \cite{Vol06}.
The comparison of $^{112}$Sn and $^{120}$Sn nuclei shown in Fig. \ref{FIG6AB} (right panel) shows that the relative difference between the E1 strengths calculated for certain energy regions, both for the QRPA and QPM cases, reduces with the increase of the excitation energies toward GDR. This behavior is expected because of the weak dependence of GDR on N/Z ratios in tin isotopes.

Furthermore, in theoretical investigations of 2$^+$ excitations in Sn isotopic chain, we find a strength clustering of quadrupole states, at low-energies, with predominantly neutron structure. 
At the same time, the proton contribution to state vectors and B(E2) strengths, located in the  energy range E$^{*}\approx$ 2-4 MeV, increases toward $^{104}$Sn and brings to more intensive proton quadrupole excitations in $^{104}$Sn there. Consequently, the $^{104}$Sn nucleus appears to be an opposing case where a change from a neutron to a proton skin occurs.
Theoretical results of B(E2) strength distributions in $^{104,120,134}$Sn isotopes are presented in Fig. \ref{FIG7}. A sizable increase of B(E2) strength at E$^{*}\approx$ 2-4 MeV is observed for the heaviest tin isotopes - $^{130}$Sn and $^{134}$Sn - studied in \cite{Tso11}.
The clustering of quadrupole states at low-energies shows a pattern similar to the PDR phenomenon. Therefore, we may consider the spectral distribution of a Pygmy Quadrupole Resonance (PQR). Correspondingly, the connection of the PQR with neutron or proton skin oscillations is demonstrated in analysis of transition densities (see in Ref.\cite{Tso11}). Similarly to the PDR, a transition from a neutron PQR to a proton PQR in $^{104}$Sn is observed for the mass region where the neutron skin reverses into a proton skin \cite{Tso11} and in Fig. \ref{FIG5AB} (left panel).

From the detailed analysis of the state vectors structure of the 2$^+$ excitations in Sn nuclei at excitation energies $E^{*}=2-4 MeV$ we find that it is dominated by neutron 2QP excitations from the valence shells. The most important proton contribution in all isotopes is due to the ${[1g_{9/2}2d_{5/2}]}_{\pi}$ 2QP component, which, however, never exceeds 5$\%$ \cite{Tso11}. This dependence is illustrated for the $[2^{+}_{2}]_{QRPA}$ states in Fig. \ref{FIG8AB} (left panel).
The figure reflects the change of the proton binding energy $\epsilon_{b}$ of the $g_{9/2}$ level when approaching the $N=Z$ limit \cite{Tso11}. 
In the neutron sector, the contributions follow closely the evolution of the shell structure as seen from Fig. \ref{FIG8AB} (left panel). In most cases the  $[2^{+}_{2}]_{QRPA}$ state vectors are dominated by re-scattering contributions related to re-orientation of the s.p. angular momenta \cite{Tso11}. 

QRPA calculations of summed B(E2) strengths in $^{104\div134}$Sn isotopes are presented in Fig. \ref{FIG8AB} (right panel). A sizable increase of B(E2) values summed at $E^{*}\approx$ 2-4 MeV is observed with increasing of $N/Z$ ratio \cite{Tso11} associated with the mass dependence of the neutron 2QP transition matrix elements and decreasing neutron binding energy $\epsilon_{b}$. 
For $N/Z<1.4$ the summed B(E2) values follow the increasing contribution of the protons, which are coupled directly by their physical charge to the electromagnetic field.

An effect of increased low-energy E2 strengths with increasing neutron number is observed as well in our QRPA calculations of lighter mass nuclei  $^{33}$Al/$^{35}$Al and $^{32}$Mg/$^{34}$Mg which are presented in Fig.  \ref{FIG9AB} (left and right panels), respectively. That mass region is of particular interest because of the well known 'island of inversion' at the N=20 shell closure \cite{Noc12}. 

Correspondingly, the connection of the low-energy 2$^+$ states with a PQR is demonstrated in the analysis of transition densities \cite{Tso11}. Results from QRPA proton and neutron quadrupole transition densities of the PQR mode in $^{124}$Sn are presented in Fig. \ref{FIG10AB} (left panel). Strong neutron oscillations at the nuclear surface play a dominant role in the PQR energy range which is in agreement with Ref. \cite{Tso11}. Similarly to the PDR, a transition from a neutron PQR to a proton PQR in $^{104}$Sn is found for the mass region where the neutron skin reverses into a proton skin \cite{Tso11}. Furthermore, QPM calculations of $B(E2)$ and $B(M1)$ transition rates of low-lying 2$^+$ states in $^{124}$Sn indicate clearly differences from known collective states and scissors modes. The results on $B(E2;g.s\rightarrow2^{+}_{i})$ transition probabilities are presented in Fig. \ref{FIG10AB} (right panel)). 

An important question to clarify is the fine structure of the observed low-energy dipole strengths. For this purpose, QPM calculations of low-energy E1 and spin-flip M1 excitations are made, in a configuration space including up to three-phonon components, in $^{138}$Ba nucleus. The results are compared to experimental data obtained from polarized photon beams \cite{Ton10}. The experimental observations and their theoretical interpretation show unambiguously the predominantly electric character of the observed low-energy dipole strength (see Fig. \ref{FIG11AB}). 

In conclusion, in systematic investigations of low-energy excitations of different multipolarity in Z=12, 13 , 50 isotopic and N=50, 82 isotonic chains a specific signals of new modes of excitations related to PDR and PQR are observed. As common feature of the structure of the PDR excited states in neutron-rich nuclei is dominated by neutron components while with the increase of the proton number, toward N=Z it transforms into a proton one. Consequently, the PDR transition strength is found directly related to the thickness of a neutron or proton skin. Its generic character is further
confirmed by related transition densities, showing that this mode
is clearly distinguishable from the conventional GDR mode.
In investigations of low-energy quadrupole states, PQR mode resembling the properties of the PDR is identified.
An especially interesting aspect is the close relation of the PQR excitations to the shell structure of the nucleus. The most convincing evidence for this feature is the disappearance of the PQR component in the double magic $^{132}$Sn nucleus. Thus, the PQR states are containing important information on the structure of the valence shells and their evolution with the nuclear mass number.

In N=82 nuclei the fine structure of the low-energy dipole strength is explained as being mostly of electric character. Even though the magnetic  contribution to the PDR is small in comparison with the electric one it should be considered of a great importance for the understanding of the spin dynamics of the nucleus in the presence of skin.


\begin{thebibliography}{}
\bibitem{Zil02} A. Zilges {\it et al.}, {\em Phys. Lett. B} {\bf 542}, 43 (2002).
\bibitem{Adr05}P. Adrich et al., {\em Phys.\ Rev.\ Lett.} {\bf 95}, 132501 (2005).
\bibitem{Vol06} S. Volz et al., {\em Nucl. Phys. A} {\bf 779}, 1 (2006).
\bibitem{Sch08} R. Schwengner et al., {\em Phys. Rev. C} {\bf 78}, 064314 (2008) and refs. therein.

\bibitem{Ton10} A. Tonchev et al., {\em Phys. Rev. Lett.} {\bf 104}, 072501 (2010).	

\bibitem{Endr10} J. Endres et al., {\em Phys. Rev. Lett.} {\bf 105}, 212503 (2010).

\bibitem{Tso04} N. Tsoneva, H. Lenske, Ch. Stoyanov,
{\em Phys.\ Lett.\ B} {\bf 586}, 213 (2004).

\bibitem{Tso08} N. Tsoneva, H. Lenske, {\em Phys. Rev. C} {\bf 77}, 024321 (2008) and refs. therein.

\bibitem{Paa07} N. Paar et al., {\em Rep. Prog. Phys.} {\bf 70}, 691 (2007) and refs therein.

\bibitem{Tso11} N. Tsoneva, H. Lenske, {\em Phys. Lett. B} {\bf 695} 174 (2011).

\bibitem{Sol76} V.G. Soloviev, {\it Theory of complex nuclei} (Oxford: Pergamon Press, 1976). 

\bibitem{Hofmann} F. Hofmann and H. Lenske, {\em Phys.~Rev.~C} {\bf 57}, 2281 (1998).

\bibitem{Vdo}A. Vdovin, V.G. Soloviev, {\em Phys. of Elem. Part. and Atom. Nucl.}, V.14, N2, 237 (1983).

\bibitem{HoKohn:64}P.Hohenberg, W. Kohn, {\em Phys. Rev.} {\bf 136}, B864 (1964).

\bibitem{KohnSham:65}W. Kohn, L.J. Sham, {\em Phys. Rev. } {\bf 140}, A1133 (1965).

\bibitem{Audi95} G.Audi, A.H. Wapstra, Nucl. Phys.~{\bf A595}, 409 (1995).
\bibitem{Gri94} M. Grinberg, Ch. Stoyanov, {\em Nucl. Phys. A} {\bf 573}, 231 (1994). 

\bibitem{Pon98} V.Yu.~Ponomarev, Ch.~Stoyanov, N.~Tsoneva, M. Grinberg, {\em Nucl. Phys. A} {\bf 635} 470 (1998).

\bibitem{Govaert98} K.\ Govaert et al. Phys.\ Rev.\ {\bf C57}, 2229 (1998).

\bibitem{Ozel} B. $\ddot{O}$zel et al., arXiv:0901.2443.

\bibitem{Kras99}A. Krasznahorkay et al., Phys. Rev. Lett. {\bf 82} 3216 (1999); A. Krasznahorkay et al., AIP {\bf 610} 751 (2002).

\bibitem{Noc12}C. Nociforo et al., Phys. Rev. C {\bf 85:044312} (2012).
\end{thebibliography}
\end{document}